\documentclass{iaus}
\usepackage{graphicx}

\title[Atmospheres and Winds of PN Central Stars] 
{Atmospheres and Winds of PN Central Stars}

\author[Kudritzki, Urbaneja \& Puls]   
{R.P. Kudritzki$^1$, M.A. Urbaneja$^1$, \and J. Puls$^2$}
\affiliation{$^1$Institute for Astronomy, University of Hawaii,
2680 Woodlawn Drive, Honolulu, HI 96822, USA \break email: 
kud@ifa.hawaii.edu\\
$^2$Institut f\"ur Astronomie und Astrophysik der Universit\"at M\"unchen,
Scheinerstr. 1, 81679 M\"unchen, Germany}

\pubyear{2006}
\volume{234}  
\pagerange{1--8}
\jname{Planetary Nebulae in our Galaxy and Beyond}
\editors{M. J. Barlow \& R. H. M\'endez, eds.}
\begin{document}

\maketitle
\begin{abstract}

The progress over the last years in modelling the atmospheres and winds of 
PN central stars is reviewed. We discuss the effect of the inclusion of the 
blanketing by millions of metal lines in NLTE on the diagnostics of 
photospheric and stellar wind lines, which can be used to determine 
stellar parameters such as effective temperature, gravity, radius, 
mass loss rate and distance. We also refer to recent work on the winds 
of massive O-type stars, which indicates that their winds are possibly 
inhomogeneous and clumped. We investigate implications from this work 
on the spectral diagnostics of PN central stars and introduce a method to
determine wind clumping factors from the relative strengths of H$_{\alpha}$
and HeII 4686. Based on new results we discuss the wind properties of CSPN.

\end{abstract}
\firstsection 

\section{Introduction: a brief history}

Since many decades model atmospheres have been a fundamental tool to 
understand the physical nature of the PN Central Stars and the 
ionization and emission of their surrounding nebulae. After the pioneering
work by Aller (1948), which revealed the importance and enormous potential
of CSPN spectroscopy, and Heap (1977, and references therein), which
provided the first quantitative spectral analyses based on model
atmospheres, the field was advanced by
M\'endez et al. (1982), Kudritzki and M\'endez (1987), M\'endez et al. (1988).
In this work, high quality spectra obtained with new 4m-class telescopes
and very efficient spectrographs and detectors were analyzed in detail
using a new generation of hydrostatic, planeparallel NLTE model 
atmospheres to determine effective temperatures, gravities and helium
abundances. This work demonstrated nicely that O-type CSPN form an
evolutionary sequence in the (log g, log T$_{eff}$)-plane and that the
gravities and temperatures determined spectroscopically could be used to
estimate stellar masses, radii, luminosities and distances by 
comparison with the prediction of post-AGB evolution and the core mass
- luminosity relationship.

While this new concept seemed compelling, there were clear indications of
quantitative deficiencies. The masses determined seemed systematically
larger than White Dwarf masses and some of the objects (such as NGC 2392)
had unrealistically high masses. The model atmospheres used, though in NLTE
and certainly state-of-the art, did not include the opacities of metal lines
and they also neglected the effects of stellar winds and spherical extension,
which were suspected to be substantial, in particular for cooler objects,
where the gravities are lower, and more massive objects, which are closer to
the Eddington limit. 

Indeed, Perinotto (1987) and Kudritzki and M\'endez (1987) stressed the
importance of stellar winds not only for the evolution of CSPN but also
for their diagnostics. With a new generation of ``unified model atmospheres'',
which included the effects of stellar winds and spherical extension, many CSPN
were re-analysed and, indeed, somewhat lower masses were found (Kudritzki
and M\'endez, 1992, Kudritzki et al., 1997). In addition, the mechanical 
momenta
of the stellar winds determined were in rough agreement with general scaling 
relations obtained from the theory of radiation driven winds and compared to
the momenta of massive O-stars, however the scatter around this relationship
was large.

The major remaining model atmosphere deficiency at this stage was the
neglect of metal line opacity, which - if included - needed to be calculated
in NLTE, certainly a formidable problem. This problem has been overcome in
recent years and a wide variety of very efficient model atmosphere codes
does exist now taking into account effects of millions of metal lines
in NLTE and the velocity fields of stellar winds together with spherical 
extension. These new codes allow for detailed studies of the UV spectra and 
a re-analysis of the optical spectrum now with the inclusion of 
line-blanketing. This is the subject of the review presented here. We will
focus on low gravity, relatively cool O-type CSPN. WR-type objects and objects
of higher gravity are discussed in other reviews of these proceedings.

\section{UV spectroscopy of CSPNs}\label{sec:uvspc}

A significant number of good UV-spectra of CSPN obtained with IUE, HST, and 
recently with FUSE are available. Many of them show the signatures of
stellar winds through broad P-Cygni profiles of resonance lines, which are
frequently used to determine terminal velocities of the stellar winds and
estimates of mass-loss rates. However, the latter are usually very uncertain,
either because the wind lines are strongly saturated or because the
ionization equlibria in the wind are uncertain and affected, for instance, by
the presence of soft X-rays and EUV radiation emitted in stellar wind shocks.

On the other hand, there are also thousands of photospheric metal lines in
the UV spectra of CSPN and their analysis provides independent means to
determine effective temperatures through photospheric ionization equilibria
such as FeIV/V. They also allow for an accurate determination of stellar
metallicity. Pauldrach et al. (2004) and Herald and Bianchi (2004a) have
carried out such studies and determined temperatures and metallicities for a
larger sample of CSPN. Compared to M\'endez et al. (1988) and Kudritzki et al. 
(1997) this work generally confirms the effective temperatures derived from
the optical line spectrum. This is important in the cases of those CSPN,
which have a much higher HeII ``Zanstra-temperature'' (the standard example
is NGC 2392). The UV work makes it clear that the high nebular ionization 
observed in these cases is not caused by a central star with an extremely high
atmospheric temperature.

The downside of the ``photospheric'' UV-work is that it does not allow for a
direct 
spectroscopic determination of stellar gravities. Thus, if not combined with
optical spectroscopy, there is no direct spectroscopic way to determine
masses, radii, and luminosities. This is only possible, if independent
assumptions about the distance are made. A beautiful example is the work by
Herald and Bianchi (2004b) of 7 LMC CSPN. Assuming a distance to the LMC,
the determination of T$_{eff}$ from the UV spectrum allows to determine radii,
luminosities and, then, with post-AGB evolution, stellar masses. Very
convincingly, the authors obtain stellar masses between 0.55 and 0.65
M$_{\odot}$.

Pauldrach et al. (2004) use a very interesting different approach. Realizing
that the theory of radiation driven winds predicts a strong dependence of
mass-loss rates and terminal velocities on stellar luminosity and stellar mass
(see Kudritzki and Puls, 2000, and references therein), they use a concept
first worked out by Kudritzki et al. (1992) to determine stellar masses and
luminosities from the observed terminal velocities and the UV-mass-loss
rates. They study the same CSPN sample as Kudritzki et al. (1997) and obtain
very similar effective temperatures. But for many objects, the masses and
luminosities are significantly different leading to the conclusion that
either the stellar wind hydrodynamics or the
core mass-luminosity relationship of post-AGB evolution, which was the basis
for the work by Kudritzki et al. (1997), are not completely accurate.

The second conclusion, if true, would have enormous repercussions for the
interpretation of post-AGB evolution. Looking critically at the results
obtained by Pauldrach et al., we note that only two of their nine objects have
masses below 0.8 M$_{\odot}$, five have masses between 1.3 and 1.4 M$_{\odot}$
just below the Chandrasekhar limit, and two are in between. For a number of
reasons that seems to be in conflict with galactic evolution and dynamics (see
Napiwotzki, 2006). We also note that in their determination of mass-loss rates
Pauldrach et al. (as Kudritzki et al.) assumed homogeneous, unclumped winds,
an assumption which might not be justified, as we will discuss later. A
re-analysis of the optical spectra of these CSPN, now using blanketed models,
can perhaps help to clarify the situation. This will also be described later.

\begin{figure}
\includegraphics[height=.357\textheight,angle=90]{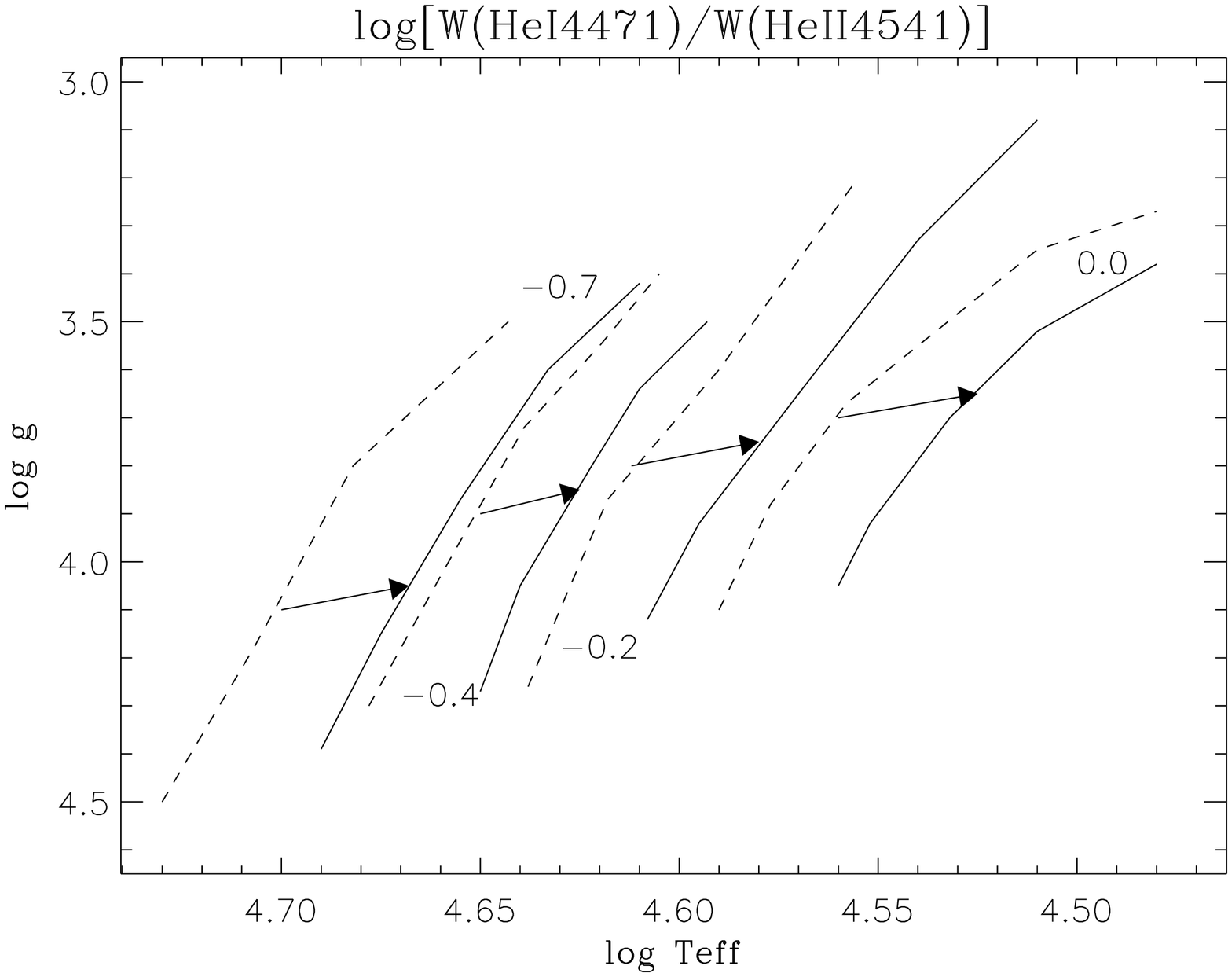} 
\includegraphics[height=.357\textheight,angle=90]{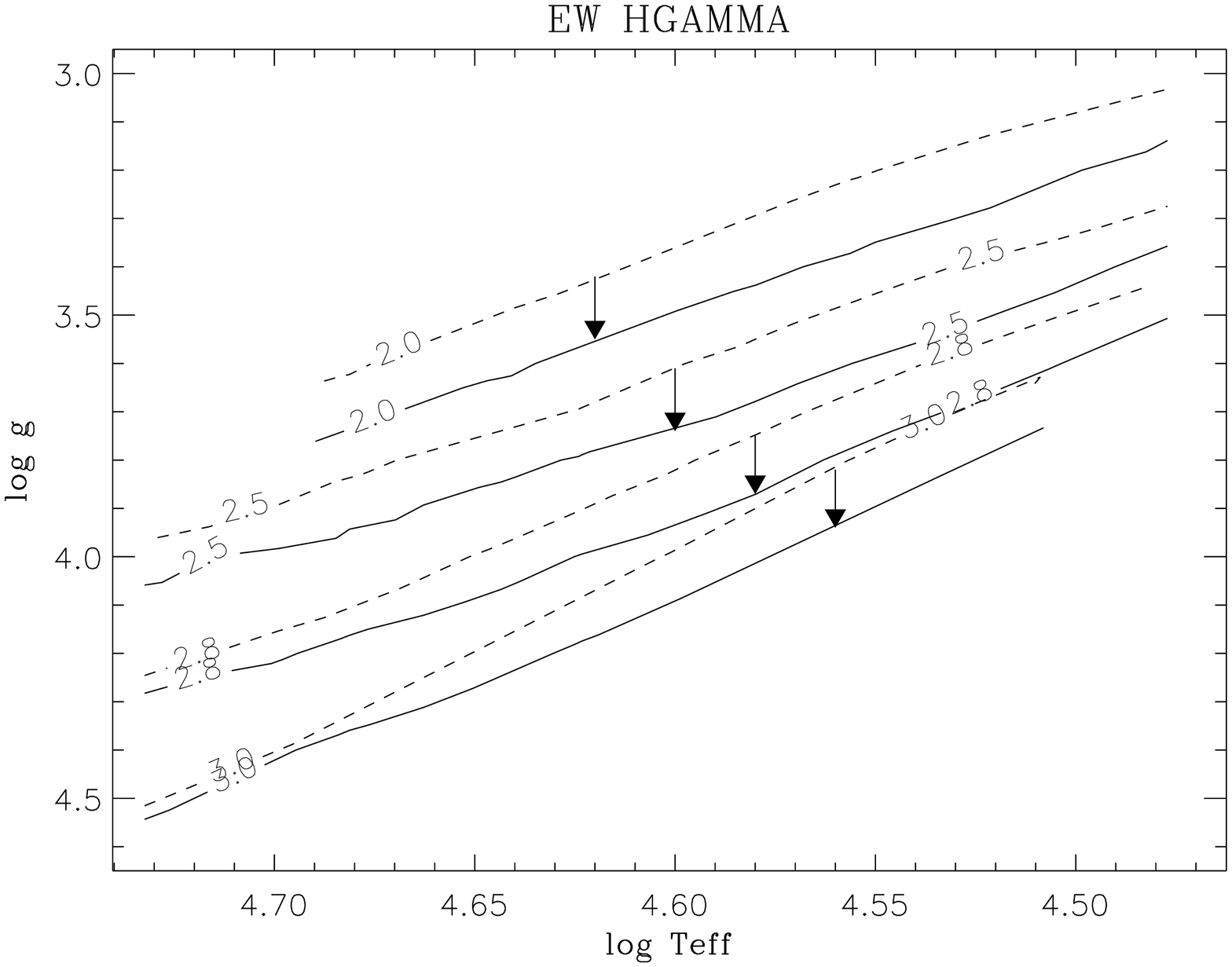} 
\includegraphics[height=.357\textheight,angle=90]{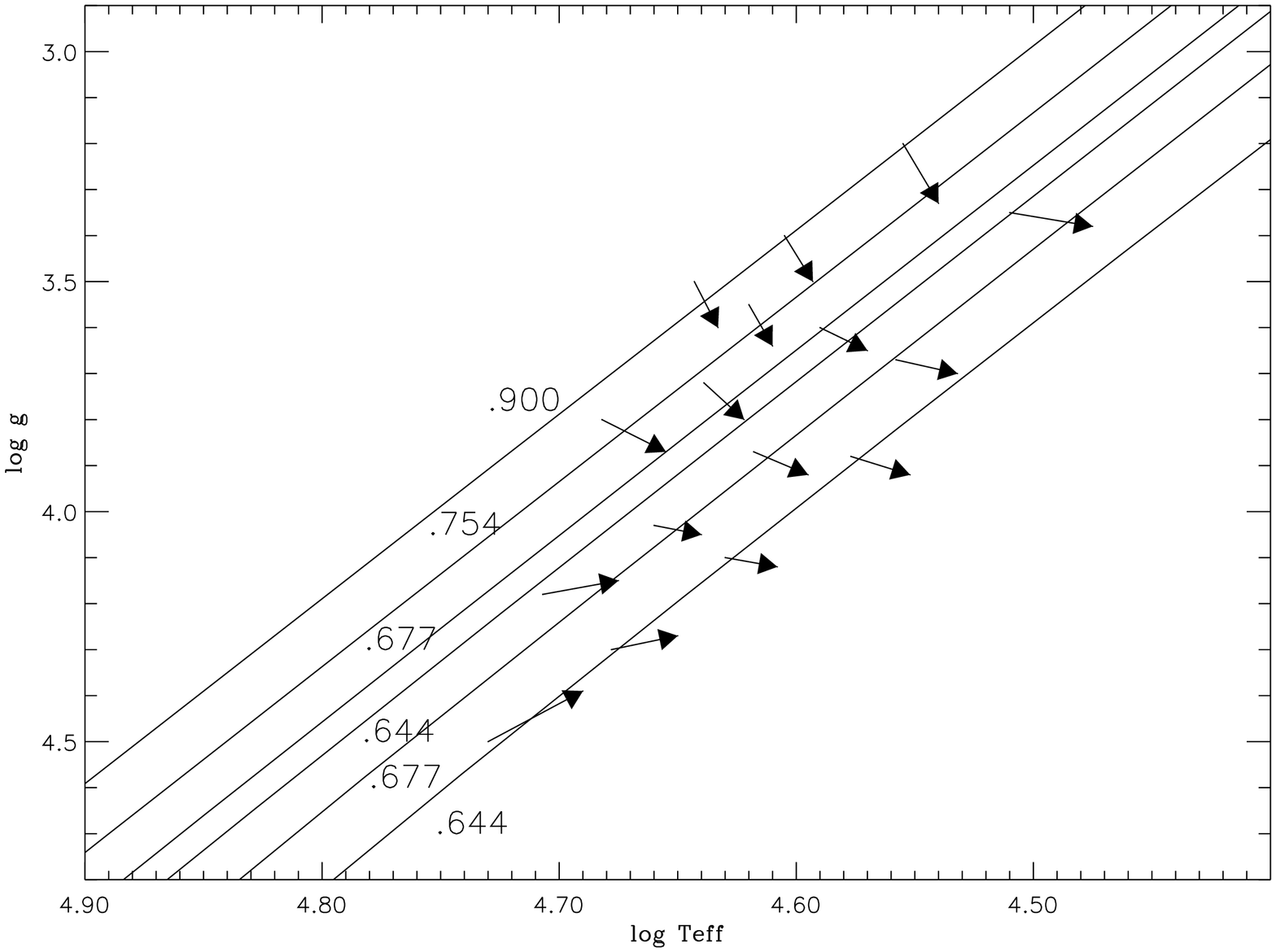}
\includegraphics[height=.357\textheight,angle=90]{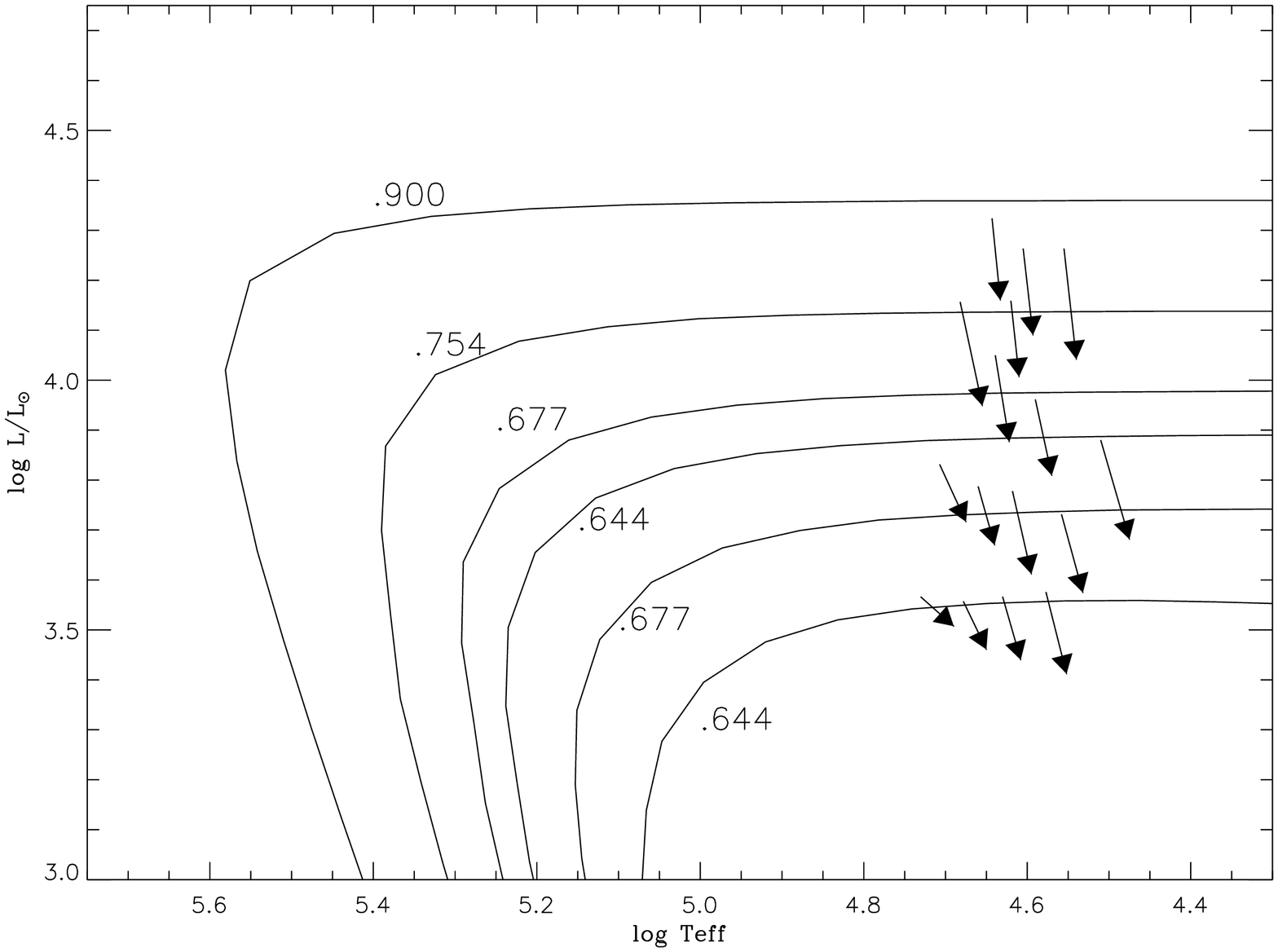} 
\caption{Upper part: isocontours of the logarithm of 
log [W$_{\lambda}$(HeI4471)/W$_{\lambda}$(HeII4542)] and
W$_{\lambda}$(H$_{\gamma}$) in the (log g, log T$_{eff}$) - plane.
Dashed isocontours are unblanketed models, solid are blanketed. Vectors
indicate the shifts caused by the effects of NLTE metal line blanketing. 
The calculations were done with the NLTE code FASTWIND (Puls et al., 2005).
Lower part: Shifts in the (log g, log T$_{eff}$) - plane (left) and
HRD (right) caused by the use of blanketed model atmospheres overplotted to
post-AGB evolutionary tracks by Vassiliades et al., 1994. The tracks are
labelled by their stellar masses.
}\label{fig1}
\end{figure}

\section{The effects of metal line blanketing}\label{sec:metlb}

The inclusion of the opacity of millions of spectral lines in NLTE has two
major effects. First, it changes the spectral energy distribution in
the UV because of strong metal line absorption in the outer atmosphere
(``line-blanketing''). However, about 50 percent of the photons absorbed are 
scattered back to the inner photosphere providing additional energy input 
and, thus, heating of the deeper photospheres. This second ``backwarming''
effect increases the continuum emission from the photosphere and
modifies ionization equilibria such as HeI/II, which are used for the
determination of T$_{eff}$. Fig.~\ref{fig1} demonstrates how the HeI/II
ionization equlibrium is shifted towards lower T$_{eff}$ because of the
backwarming effect. At the same time, the pressure-broadened wings of the
Balmer lines (the standard diagnostic for log g) become weaker, because
the millions of metal lines increase the radiative acceleration g$_{rad}$
and decrease the effective gravity g$_{eff}$ = g - g$_{rad}$. As a result
higher gravities are needed to fit the Balmer lines (see Fig.~\ref{fig1})
in addition to the lower temperatures obtained from the helium ionization
equlibrium. In summary, the use of blanketed models leads to systematic
shifts in the (log g, log T$_{eff}$) - plane, which if compared with
post-AGB evolutionary tracks result in systematically lowering  CSPN masses, 
radii, luminosities and distances. Note that the presence
of dense stellar wind envelopes increases the effects of backwarming and
introduce an additional dependence on mass-loss rates (see Sellmaier et al.,
1993, Repolust et al., 2004).

The combined effects of line blanketing and backwarming affect also the
ionizing fluxes. Amazingly, for the ionization of hydrogen the changes are
very small as the effects of blanketing and backwarming balance each other.
However, the ionization of ions with absorption edges shorter than the one
for hydrogen is significantly affected (see Kudritzki, 2002, Martins et al.,
2005).

\begin{figure}
\includegraphics[height=.357\textheight,angle=90]{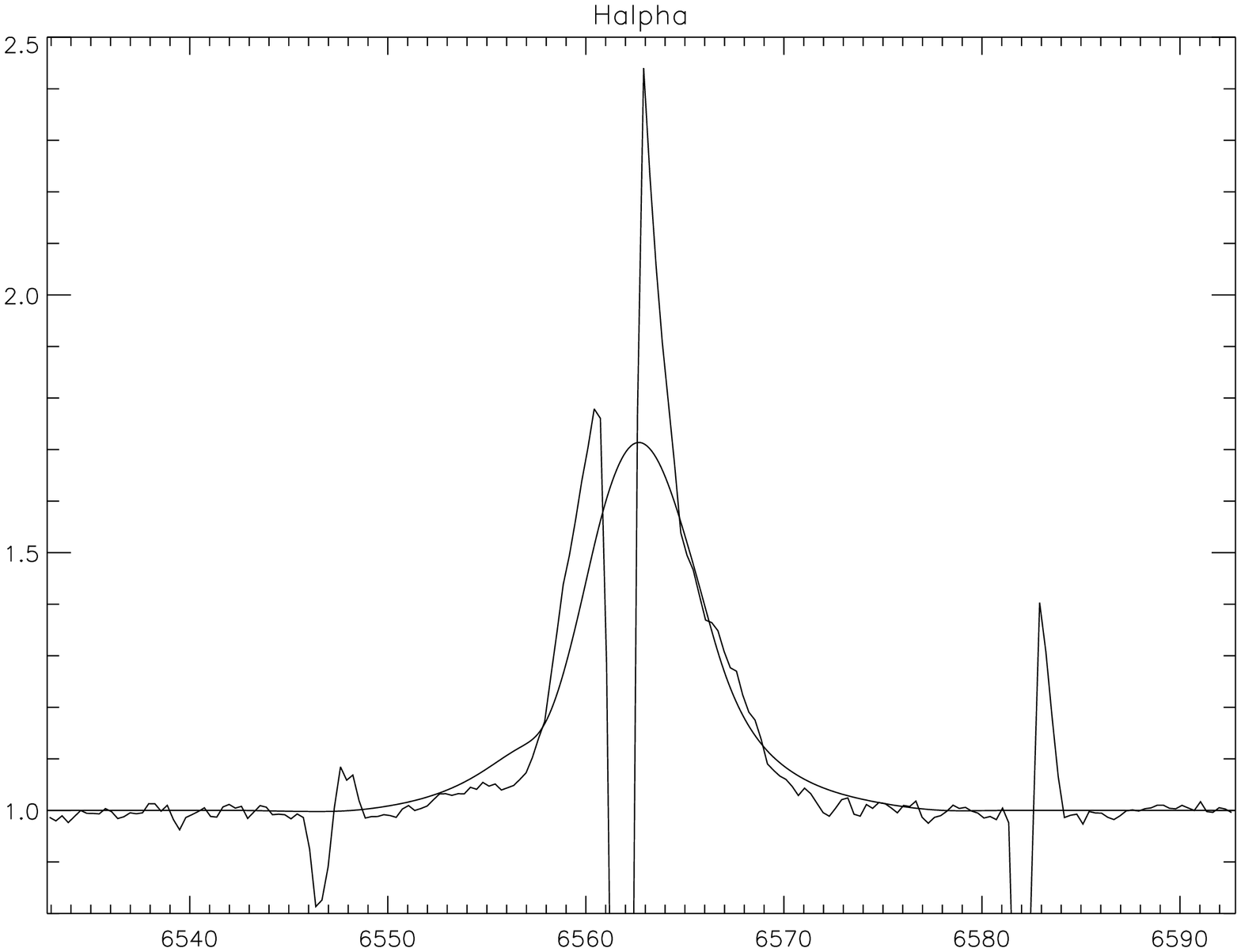} 
\includegraphics[height=.357\textheight,angle=90]{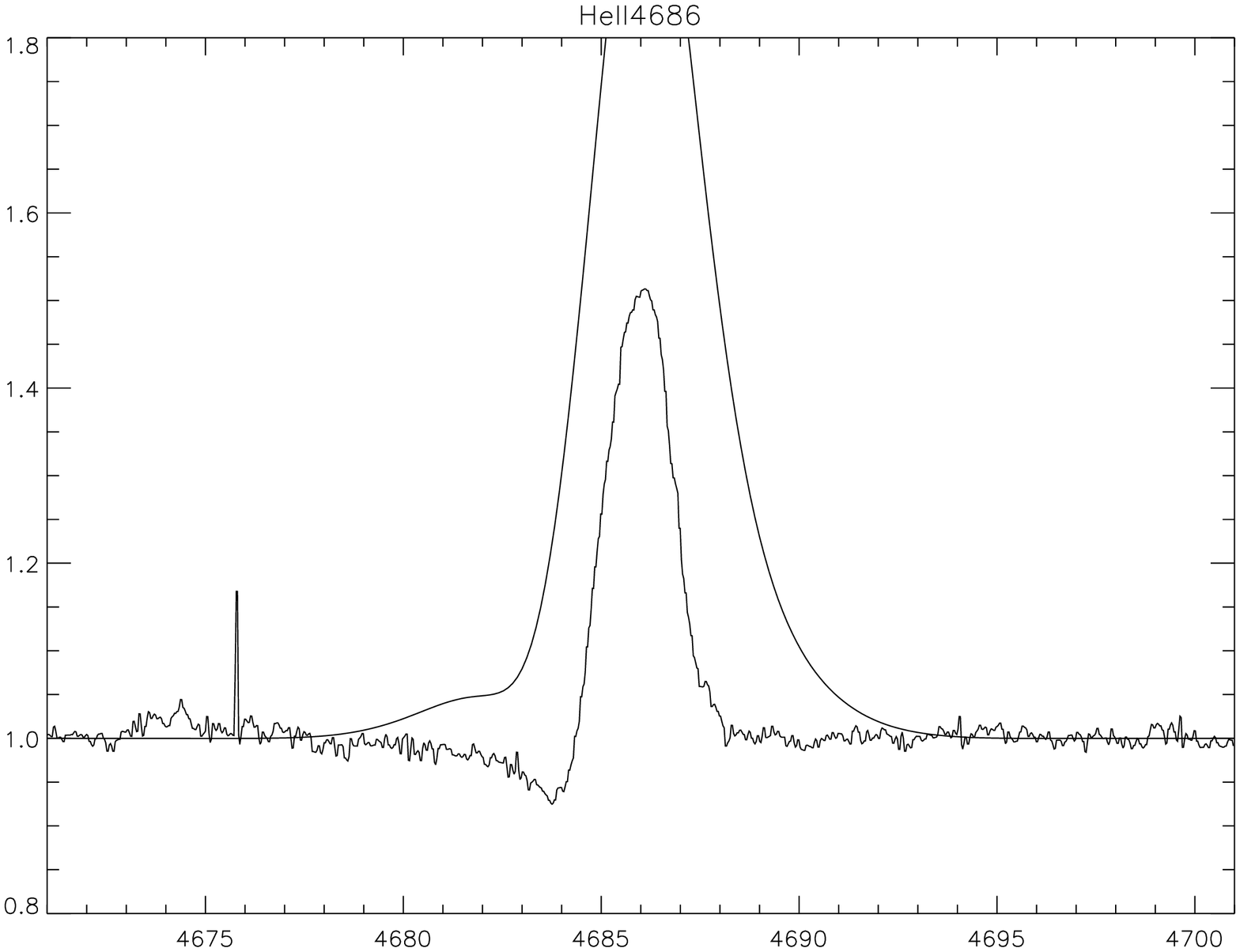} 
\includegraphics[height=.357\textheight,angle=90]{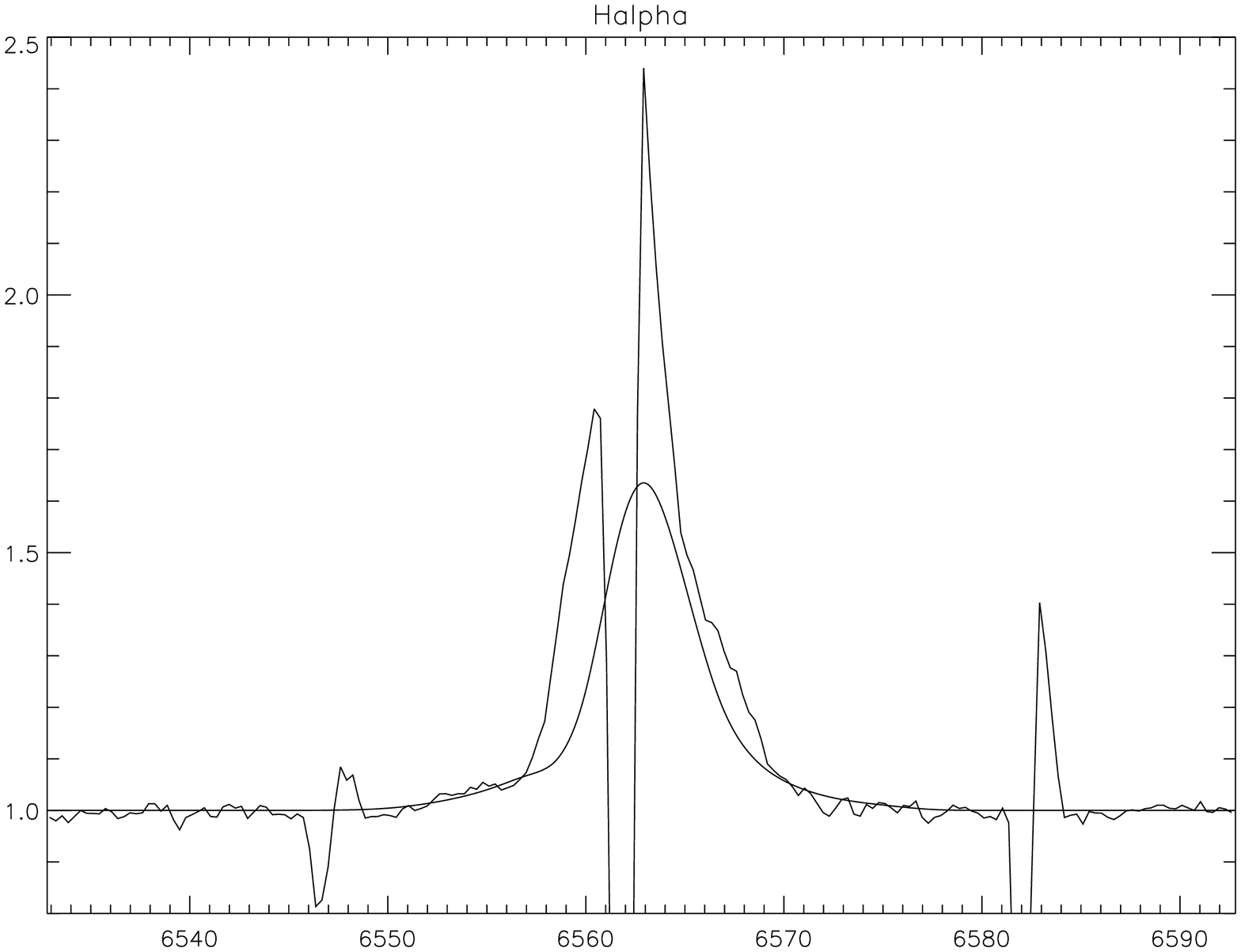} 
\includegraphics[height=.357\textheight,angle=90]{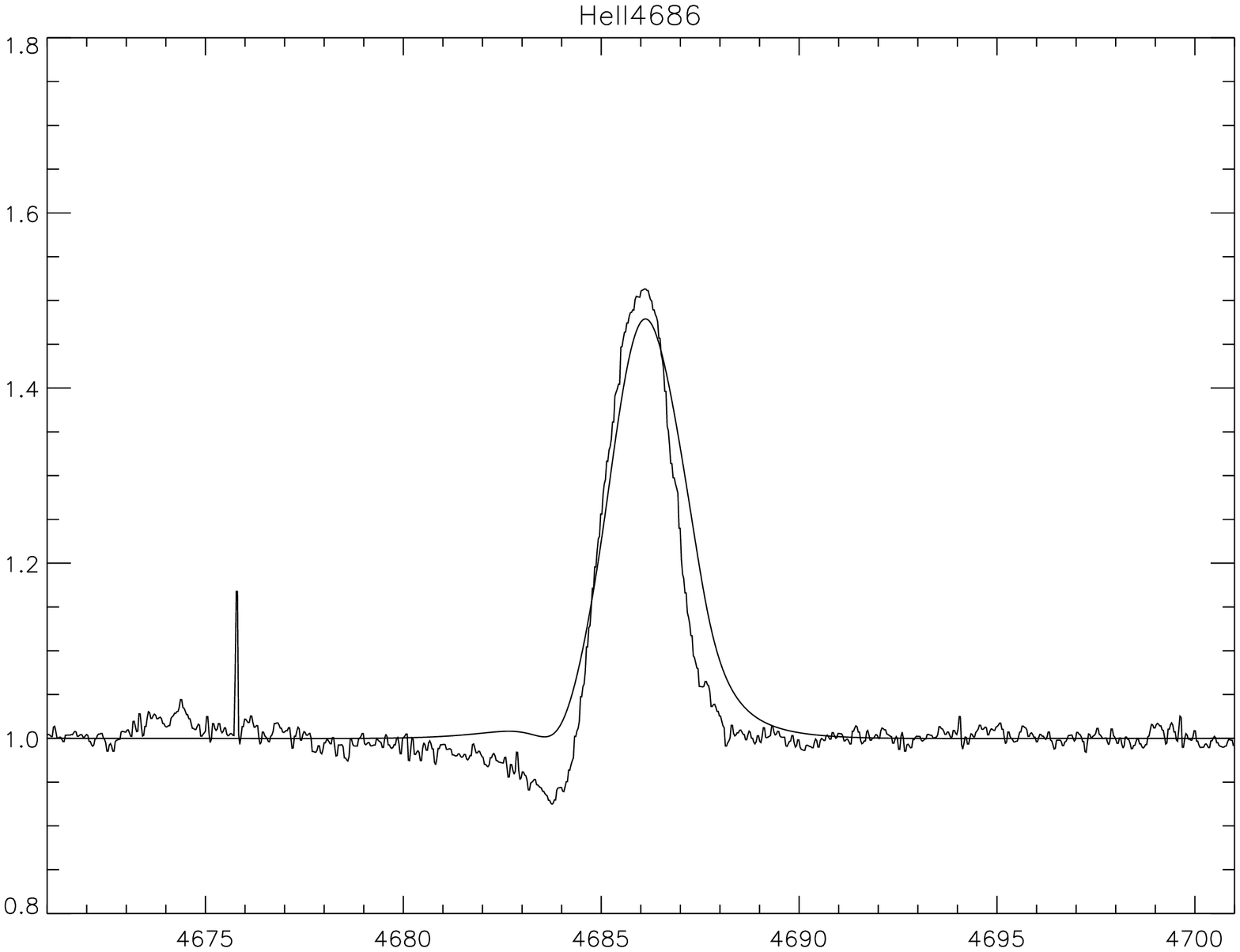} 
\caption{Diagnostics of stellar the stellar wind emission lines H$_{\alpha}$
(left) and HeII 4686 (right) of IC 418. In the top row $f_{cl} = 1$ is
adopted and in the bottom we use $f_{cl} = 50$. Note that for H$_{\alpha}$
nebular lines have been (imperfectly) subtracted.
}\label{fig2}
\end{figure}

\begin{figure}
\includegraphics[height=.357\textheight,angle=90]{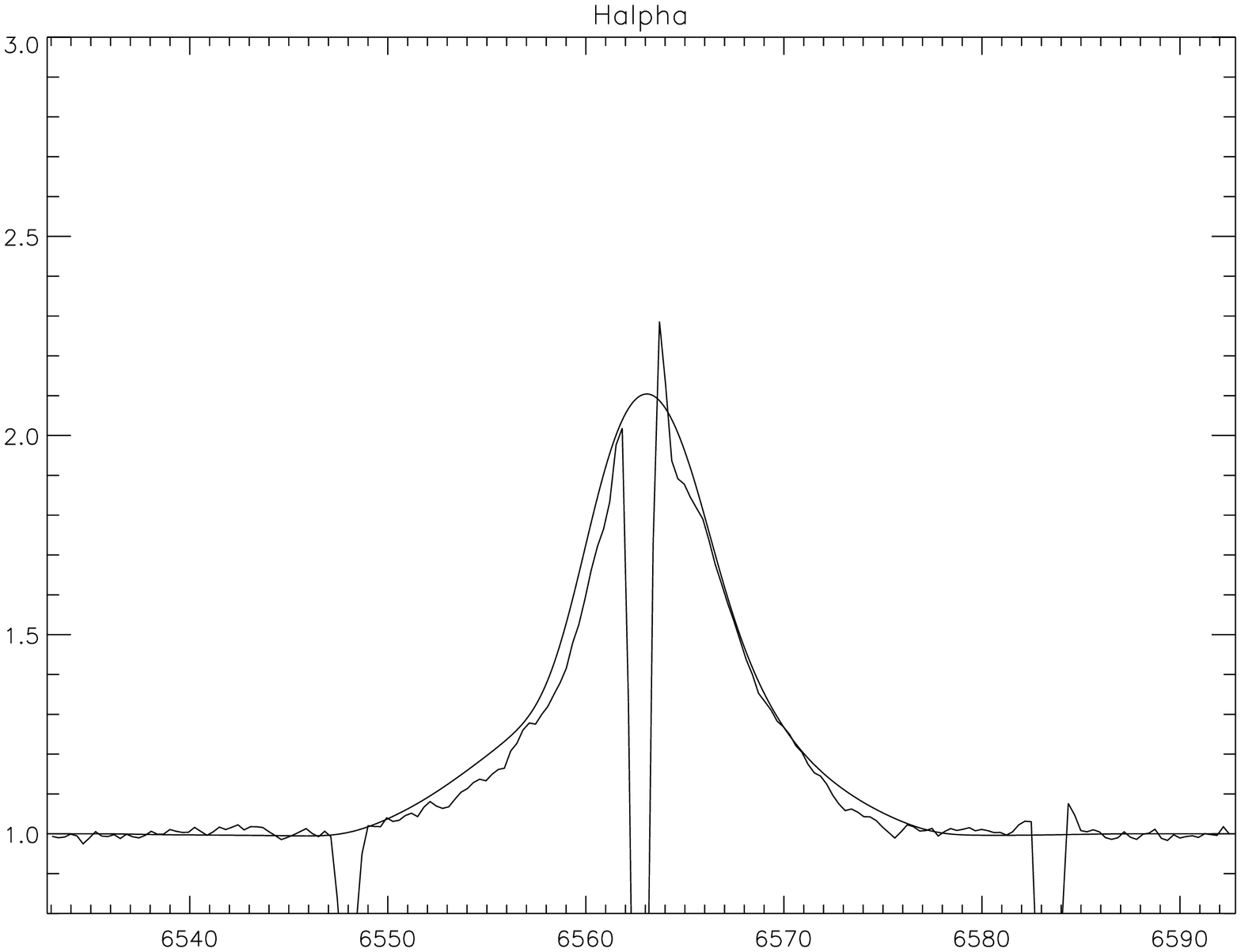} 
\includegraphics[height=.357\textheight,angle=90]{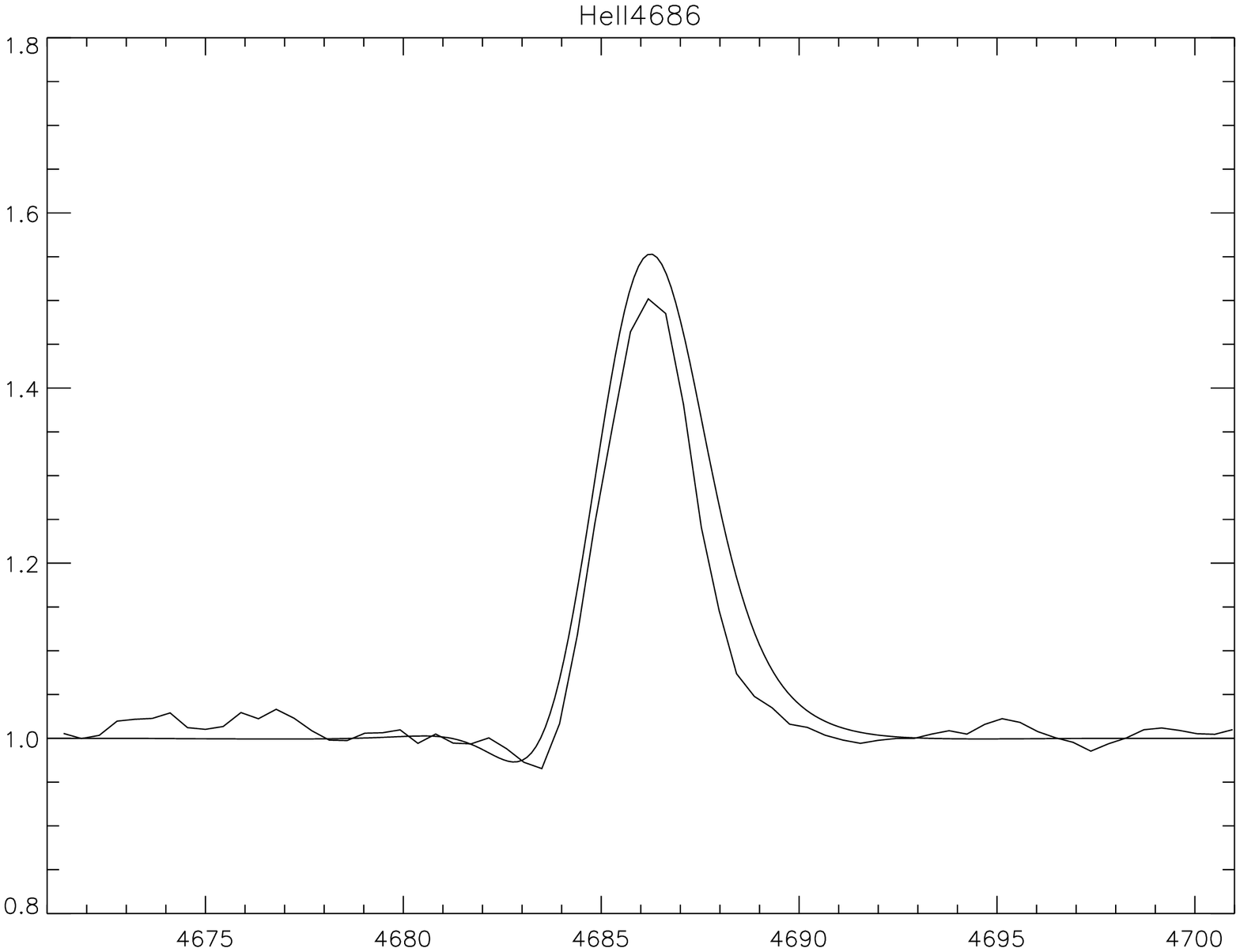} 
\caption{Diagnostics of stellar the stellar wind emission lines H$_{\alpha}$
(left) and HeII 4686 (right) of He 2-108. $f_{cl} = 1$ is adopted for the fit.
Same as Fig 2 for the nebular lines.
}\label{fig3}
\end{figure}

\section{Detailed analysis of optical spectra and the effects of wind clumping}\label{sec:detan}

The significant effects caused by NLTE line-blanketing make it worthwhile to
re-analyse the optical spectra of the sample studied by Kudritzki et al.
(1997). This will also allow for a comparison with the UV-study carried out
by Pauldrach et al. (2004) discussed above. For our analysis we use the NLTE
code FASTWIND (Puls et al., 2005), which includes the effects of NLTE metal
line opacities, stellar winds, and spherical extension. The strategy for the
analysis is identical to Kudritzki et al. (1997) (see also Repolust et al.,
2004, for more recent work). T$_{\rm eff}$ and helium abundance are obtained
from a fit of the HeI and HeII lines, while the gravity is determined from
the higher Balmer lines. H$_{\alpha}$ as the strongest optical hydrogen line
is formed in the stellar wind and, thus, used to constrain the mass-loss
rate. The terminal velocity follows from fits of the UV P-Cygni lines.  

While H$_{\alpha}$ is, in principle, a perfect tool to measure mass-loss 
rates (see Kudritzki and Puls, 2000, Kudritzki, 2006, for discussion and 
references), the results might be affected by stellar wind clumping. It has
been known since long that line driven winds are intrinsically unstable
(Owocki et al., 1988, 2004). This might lead to inhomogeneous, clumped winds
such as described by Owocki and Runacres (2002) with regions of enhanced
density $\rho_{cl}$ and regions, where the density is much lower. In a very
simple description, introducing clumping factors $f_{cl}$ similar as in
PN diagnostics, the relationship between the average density of the stellar
wind flow $\rho_{av}$ and the density in the clumps is then given by
$\rho_{cl} = \rho_{av} f_{cl}$. The  same relationship holds for the
occupation numbers $n_{i}$ of ions.  

Line opacities $\kappa$ depend on density through $\kappa \propto n_{i}
\propto \rho^{x}$ and for very small, optically thin clumps the avarage 
optical line depth in the wind is given by 
$\tau_{av} \propto n^{av}_{i} \propto n^{cl}_{i} f^{-1} \propto
\rho_{av}^{x} f^{x-1}$. For a dominating ionization stage we have $x = 1$ 
and the clumping along the line of sight cancels and does not affect the 
diagnostics. However, bound hydrogen is a minor ionization stage in hot 
stars depending on recombination from ionized hydrogen 
with $n_{i}(H) \propto n_{E}n_{P} \propto \rho^{2}$. Thus, 
if $f_{cl}$ is significantly larger than one, the H$_{\alpha}$
mass-loss rate diagnostic is systematically affected and we have
$\dot{M}(H_{\alpha}) = \dot{M}(true) f_{cl}^{1/2}$, following from the fact
that $\dot{M}(true) \propto \rho_{av}.$

The spectral diagnostics of clumping in CSPN is difficult. In principle, 
it requires the comparison of lines with different exponents x in the 
density dependence of their opacities. In WR-type CSPN with very dense winds 
and very strong wind emission lines (see these proceedings or Hamann et al.,
2001) incoherent electron scattering produces wide emission wings, the 
strength of which goes with $x \sim 1$. Clumping factors of the order of ten
to twenty were found. This technique does not work for O-type CSPN, as their winds
have much lower density. Also the UV P-Cygni lines of dominating ions
provide usually little help, as these lines are mostly saturated and the
ionization equlibria are uncertain. However, in most recent work on massive
O-stars using FUSE and Copernicus spectra the PV resonance line at 1118 and
1128 \AA \ has been used as an indicator of clumping. The advantage of PV is
the low cosmic abundance so that the line is completely unsaturated even when
in a dominating ionization stage. Substantial 
clumping was found (Hillier et al., 2003,
Bouret et al., 2005, Fullerton et al., 2006). Unfortunately, only a
few useful FUSE spectra are available for our sample. We have, therefore,
applied a different technique to constrain clumping, at least for a subset
of our objects.

For cool O-type CSPN with T$_{\rm eff} \le 37,000$ K HeII is a dominant 
ionization
stage. That means for objects with strong winds and HeII 4686 in emission
and formed in the wind this line should have a density dependence close to 
$x =1 $. Its relative strenght to H$_{\alpha}$ should allow to constrain 
f$_{cl}$.

In the following we present the results of this new work. For lack of
space we do not show typical example of the fits of lines which constrain 
T$_{\rm eff}$, log g, and the helium abundance and refer to Kudritzki et al.
(1997). Fig.~\ref{fig2} demonstrates in our most extreme case how $f_{cl}$
is constrained. $f_{cl} = 1$ leads to by far too strong emission of HeII 4686
relative to H$_{\alpha}$. A very large value of $f_{cl} = 50$, however, 
improves the fit significantly. Fig.~\ref{fig3} gives an example 
of the other extreme where a homogeneous wind with $f_{cl} = 1$ results in
a very satisfactory fit. A summary of all results is given in 
Tab.~\ref{tab1}. Note that the last column indicates the three cases where
we were able to constrain $f_{cl}$. In the other cases, either
nebular emission did not allow for a determination (Tc 1) or the objects
were too hot for the method to be applicable. 

  \begin{table}
  \caption{Stellar parameters of CSPN analyzed}
  \label{tab1}
  \begin{center}
    \begin{tabular}{l c c c c c c c c c c c}
      \hline
 object & T$_{eff}$ & log g & He ab. & R/R$_{\odot}$ & log L/L$_{\odot}$ &
M/M$_{\odot}$ & d & $\dot{M}$ & $v_{\infty}$ & $f_{cl}$ & det. \\
      \hline
    & 10$^{3}$K & cgs &  &  &  &  & kpc & log M$_{\odot}$/yr & km/s &  & \\     
      \hline
  He 2-131 & 32 & 3.2 & .33 & 3.5 & 4.07 & .71 & 3.3 & -6.88 & 400  & 8  & y \\
  Tc 1     & 34 & 3.2 & .09 & 3.8 & 4.23 & .81 & 4.4 & -7.46 & 900  & 1  & n \\
  He 2-108 & 34 & 3.4 & .09 & 2.6 & 3.92 & .63 & 5.8 & -6.85 & 700  & 1  & y \\
  IC 418   & 36 & 3.2 & .17 & 4.0 & 4.38 & .92 & 2.7 & -7.43 & 700  & 50 & y \\
  IC 4593  & 40 & 3.6 & .09 & 2.2 & 4.05 & .70 & 3.5 & -7.36 & 900  & 4  & n \\
  NGC 2392 & 44 & 3.6 & .23 & 2.4 & 4.30 & .86 & 2.8 & -7.32 & 400  & 1  & n \\
  NGC 6826 & 46 & 3.8 & .09 & 1.8 & 4.11 & .74 & 2.6 & -7.10 & 1200 & 4  & n \\
  IC 4637  & 52 & 4.2 & .09 & 1.0 & 3.85 & .62 & 1.3 & -7.91 & 1500 & 4  & n \\
  NGC 3242 & 75 & 4.8 & .09 & 0.5 & 3.89 & .63 & 1.8 & -8.08 & 2300 & 4  & n \\
\end{tabular}
\end{center}
\end{table}

\begin{figure}
\includegraphics[height=.357\textheight,angle=90]{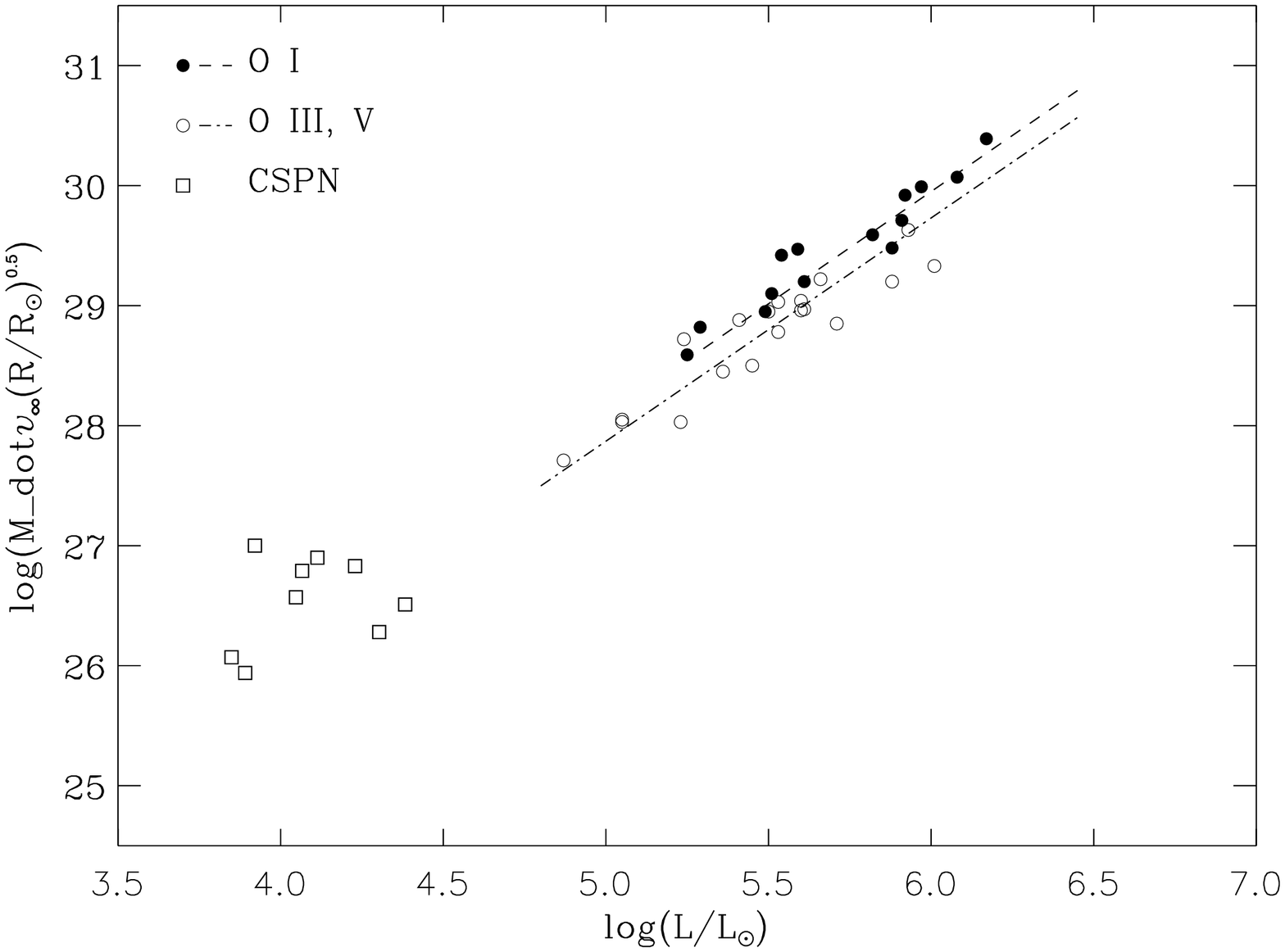} 
\includegraphics[height=.357\textheight,angle=90]{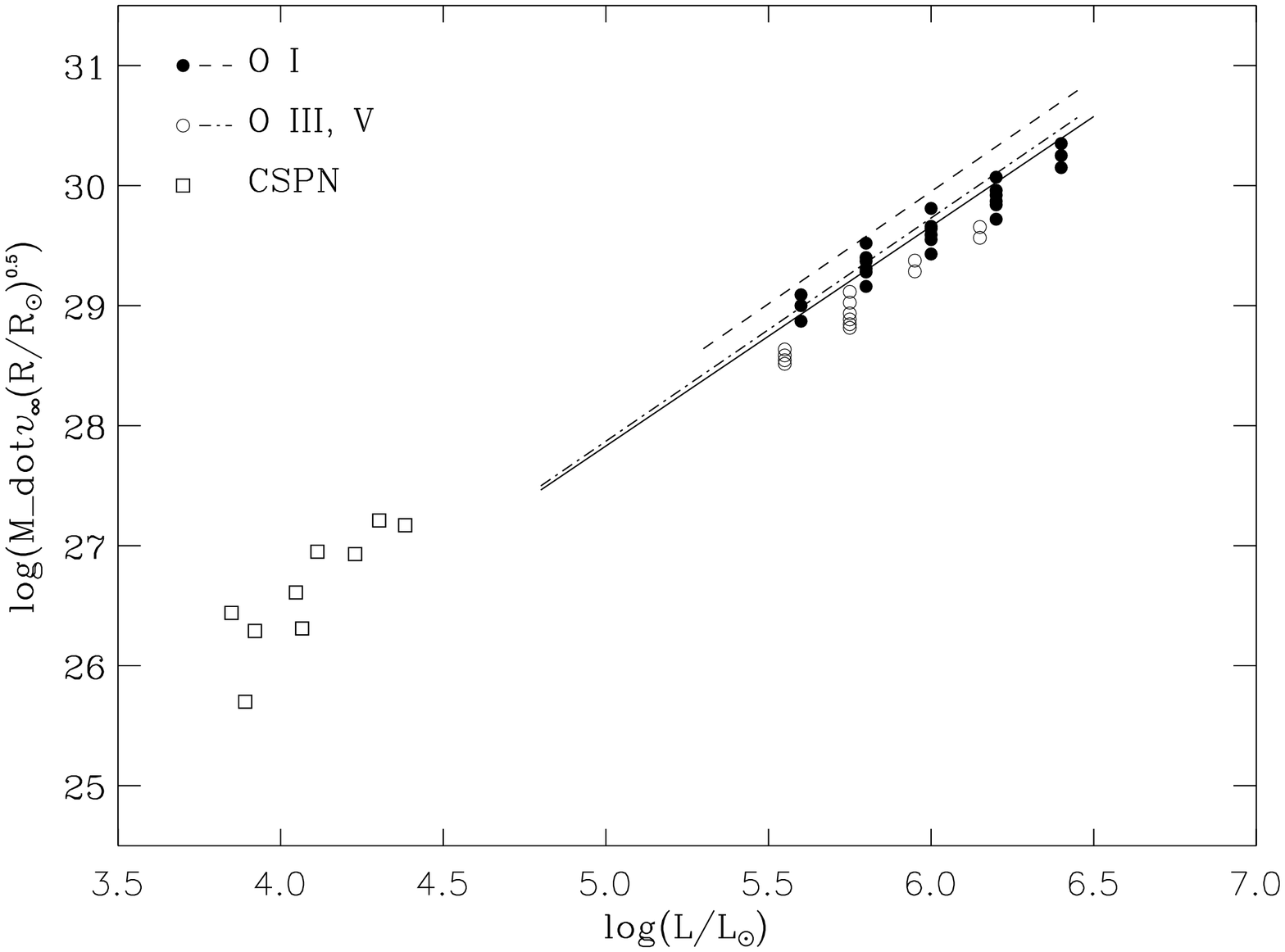}
\caption{Left: Observed CSPN stellar wind momenta (this work) compared with those of
massive O-stars (Repolust et al., 2004, Markova et al., 2004). Right:
Calculated stellar wind momenta for the CSPN of this work and massive
O-stars using the theory of line driven winds as developed by Kudritzki
(2002). Symbols refore to model calculations. The dashed lines are the
regression curves obtained from the observations of massive O-supergiants
and dwarfs. The solid line represents the theoretical approach by Vink et
al., 2000.
}\label{fig4}
\end{figure}

\section{Discussion and future work}\label{sec:disfu}

Comparing our results to Pauldrach et al. (2004) we find reasonable agreement
for the effective temperatures for all cases (except He 2-108, which we find
to be 17 percent cooler). This is very satisfying given
the fact that different spectroscopic techniques were used for the
determination of T$_{eff}$. However, in all but one case (He 2-131) the
gravities obtained through our fitting of the Balmer lines are substantially
lower. If one corrects for the temperature dependence of the Balmer lines
(see Fig.~\ref{fig1}) and the slightly different temperatures obtained, our
gravities are on average 0.3 dex smaller. The reason is clearly the
completely different approach used to determine gravities, as explained in
section 2. Pauldrach et al. rely on the hydrodynamic
simulation of stellar winds, whereas we have have used the classical
spectroscopic concept of Balmer line fitting. This will need further
investigation.

For the determination of masses, radii, luminosities, and distances we have
used again the classical approach of using the post-AGB core mass -
luminosity relationship. Compared to Kudritzki et al. (1997) our masses are
generally smaller in agreement with what we expect from section 3, however
some of them (IC 418, NGC 2392) are still uncomfortably high. 

The mass loss rates determined are uncertain because of the effects of
stellar wind clumping. However, if we take into account that mass-loss rates
determined with H$_{\alpha}$ scale with the stellar radius adopted as
$\dot{M} \propto R^{3/2}$ (Kudritzki and Puls, 2000) 
and compare with Pauldrach et al. we find
agreement within a factor of two except for IC 418. Fig.~\ref{fig4} shows
the CSPN wind momenta of our study compared to massive O-stars. There are
two ways to interprete this plot. One is that CSPN form a
convincing extension of the wind momentum - luminosity relationship of
massive O-stars towards lower luminosities. Another one is that within the
luminosity range of CSPN alone there is no clear relationship between wind
momentum and luminosity (see also Tinkler and Lamers, 2002). Whether this is
because of the uncertainties of mass-loss rate diagnostics or of the
luminosity determinations, or both, will need further investigation.

It is interesting to compare this result with stellar wind models obtained
with the theory of line driven winds. This is also done in Fig.~\ref{fig4}, 
where
we use the method by Kudritzki (2002) to calculate wind momenta for the CSPN
of Tab.~\ref{tab1}. With a few exceptions the theoretical momenta are in the
right ballpark.

Several steps need to be undertaken for future work. The first is to extend
the work presented here and to re-analyse the UV-spectra of the sample and
to compare in detail to see whether or not the
results obtained from the two spectral windows are compatible, in this way
addressing the original point made by Pauldrach et al. (2004). Taking into
account the effects of clumping through the diagnostics of the PV and similar
lines will be crucial. 

A very important issue is the compatibility of stellar wind hydrodynamics
with the stellar parameters derived by our method. While the wind momenta
seem to agree within the observational errors, the terminal velocities
calculated are too small in many cases for the parameters obtained by us.
Whether this is a deficiency of stellar wind hydrodynamics or of the Balmer
line diagnostics used, remains to be investigated. We note that one
modification to be made in the wind hydrodynamics is the inclusion of
clumping factors. It will be interesting to see whether this will help to
resolve the discrepancy.

\end{document}